\pdfoutput=1
\documentclass[conference]{IEEEtran}
\usepackage{hyperref}
\hypersetup{
pdfauthor=Havemann; Heinz; Struck; Gl\"aser,
pdftitle=Havemann et al. (2010): Identification of Overlapping Communities by Locally Calculating Community-Changing Resolution Levels,
colorlinks=true,
linkcolor=black,
anchorcolor=blue,
citecolor=blue,
urlcolor=blue,
menucolor=black
}

\usepackage{cite}
\ifCLASSINFOpdf
   \usepackage[pdftex]{graphicx}
\else
\fi
\usepackage[cmex10]{amsmath}
\usepackage{algorithmic}
\usepackage{array}
\usepackage{url}
\begin{document}
\title{Identification of Overlapping Communities \\by Locally Calculating Community-Changing Resolution Levels}
%Identification of overlapping communities by locally calculating community-changing resolution levels}

\author{\IEEEauthorblockN{\textsc{Frank Havemann} \\\textsc{Michael Heinz}\\\textsc{Alexander Struck}}
\IEEEauthorblockA{Institut f\"ur Bibliotheks- und Informationswissenschaft\\
Humboldt-Universit\"at zu Berlin\\
Unter den Linden 6\\
10099 Berlin\\
Germany%\\
}
\and
\IEEEauthorblockN{\textsc{Jochen Gl\"aser}}
\IEEEauthorblockA{Zentrum f\"ur Technologie und Gesellschaft\\
Technische Universit\"at Berlin\\
Hardenbergstr. 36A\\
10623 Berlin\\
Germany%\\
}
}


% for over three affiliations, karator if they all won't fit within the width
%Telephone: (800) 555--1212, Fax: (888) 555--1212}

%\IEEEspecialpapernotice{2010, August 5}

\maketitle

\begin{abstract}
\boldmath
The identification of thematic structures in networks of bibliographically or lexically coupled papers is hindered by the fact that most publications address more than one theme, which in turn means that themes overlap in publications.
An algorithm for the detection of overlapping natural communities in networks was proposed by Lancichinetti, Fortunato, and Kertesz~(LFK) last year~\cite{lancichinetti2009detecting}. The LFK algorithm constructs natural communities of (in principle) all nodes of a graph by maximising the local fitness of communities. The authors define fitness as the ratio of the number of internal links to the number of all links of the nodes of a community but the denominator of the ratio is raised to the power of $\alpha$. This parameter can be interpreted as the resolution at which natural communities are determined. The resulting communites can, and are due to the constructing approach likely to, overlap. The generation of communities can easily be repeated for many values of $\alpha$; thus allowing different views on the network at different resolutions.
We implemented the main idea of the LFK algorithm---to search for natural communities of each node of a network---in a different way. We start with a value of the resolution parameter that is high enough for each node to be its own natural community. When the resolution is reduced, each node acquires other nodes as members of its natural community, i.e. natural communities grow. For each community found at a certain $\alpha$ value we calculate the next lower $\alpha$ where a node is added. After adding a node to a community of seed node $k$ we check whether the natural community of node $k$ is also the natural community of a node that we have already analysed. If this is the case, we can stop analysing node $k$.
We tested our algorithm on a small benchmark graph and on a network of about 500 papers in information science weighted with the Salton index of bibliographic coupling. In our tests, this approach results in characteristic ranges of $\alpha$ where a large resolution change does not lead to a growth of the natural community. Such results were also obtained by applying the LFK algorithm but since we determine communities for all resolution values in one run, our approach is faster than the original LFK approach.\footnote{The results presented were also shown on a poster with the title \textit{A local algorithm to get overlapping communities at all resolution levels in one run} at ASONAM conference, Odense, Denmark, August 2010.} 

\end{abstract}


% \fi
\IEEEpeerreviewmaketitle


\section{Introduction}
Many real-world networks consist of substructures that overlap because nodes are members of more than one substructure. Networks of scientific papers are a case in point. Thematic structures such as common topics, approaches, or methods are not disjunct. It is the rule rather than the exception that a paper addresses more than one topic. 

Hard clustering is inadequate for the investigation of real-world networks with such overlapping substructures. Instead, methods are required that allow nodes to be members of more than one community in the network. During the last years a number of algorithms for detecting overlapping communities (or modules) in graphs have been developed and tested. One approach starts from hard clusters obtained by any clustering method and assigns the nodes at the borders of clusters to several neighbouring modules~\cite{baumes2005efficient,wang2009adjusting}. In another approach links are clustered into disjoint modules and nodes are members of all modules their links belong to \cite{ahn2009link,evans2009edge}. Our paper is based on a third approach that constructs  natural communities of all nodes which can overlap each other \cite{lancichinetti2009detecting}.

In our search for methods that model scientific specialties as networks of journal papers and enable the identification of thematic structures in those networks, we applied the algorithm developed by Lancichinetti, Fortunas, and Kertesz~\cite{lancichinetti2009detecting}. This LFK algorithm is well suited to our problem because it identifies not only overlapping communities but also a hierarchical structure of a graph if there is any. Since we assume that thematic structures are of varying scope and that some of the smaller themes might be completely contained in larger ones, an algorithm that detects both overlaps and hierarchies is essential. 

The main assumption of the LFK algorithm is that every node has its own natural community. In our context this approach can be interpreted as the construction of a thematic environment from the 'scientific perspective` of the seed paper. This idea is not only attractive from a conceptual point of view---the borders of topics are explored by a local algorithm i.e. independently from papers located far away from the seed paper---but also for services leading users of bibliographic databases from one relevant paper to thematically similar ones.

The essence of the LFK algorithm is that independently constructed natural communities of nodes can overlap. In accordance with the locality of their approach Lancichinetti, Fortunas, and Kertesz evaluate the fitness of modules of nodes with a function that uses only local information. It is based on the assumption that a community should have more internal than external links. The fitness function is defined as the ratio of the sum of internal degrees to the sum of all degrees of nodes in a module $G$. The denominator is taken to the power of $\alpha$, the resolution parameter:

\begin{equation}
 f(G, \alpha) = \frac{k_{in}(G)}{(k_{in}(G) + k_{out}(G))^\alpha}.
\end{equation} 

For each node a \textit{natural community} $G$ is constructed by including the neighbour that produces the highest fitness gain. Then the fitness gain of each node in $G$ is recalculated. If it is negative remove this node from $G$. The community is complete if including any neighbour brings no fitness gain.

The authors conclude \cite[p. 6]{lancichinetti2009detecting}: ``By varying the resolution parameter one explores the whole hierarchy of covers of the graph, from the entire network down to the single nodes, leading to the most complete information on the community structure of the network.''

Since the LFK algorithm constructs natural communities of all nodes of a graph and has to be repeated for each value of the resolution parameter within the interval of interest, applying it to larger networks is time-consuming.

Acknowledging this, the authors proposed several ways in which their algorithm could be optimised. They tested an implementation that starts from a random node and after construction of its community switches to the next random node outside this community until the whole graph is covered~(we denote this version of the algorithm by \textit{random LFK}).  Lancichinetti, Fortunas, and Kertesz also proposed to use communities found at one level of resolution as starting points for the next lower level because at lower resolution a community cannot be smaller than at higher level. 

We implemented the main idea of the LFK algorithm---to search for natural communities of each node of a network---in a different way. For some sufficiently high value of the resolution parameter alpha each node is a single, i.e. it is its own natural community. Lowering the resolution makes the single nodes include `companions' because this increases the community's fitness function. The inclusion of nodes makes the natural community of each node grow. For each community found at some alpha we look for the next lower alpha at which new members are acquired. Whenever a node is added to a natural community of seed node $k$ we check whether the natural community of node $k$ is fully contained by the natural community of any other node. If this is the case, we can stop analysing node $k$. This way, we  \textbf{m}erge (completely) \textbf{o}verlapping \textbf{n}atural \textbf{c}ommunities. Therefore we choose the acronym MONC for our algorithm. 

Since we determine communities for all resolution values in one run our algorithm is faster than the original LFK algorithm. Both algorithms are different implementations of the idea of growing natural communities of nodes, i.e. they are not totally equivalent. We discuss the differences between the two algorithms in the following section.%
\footnote{In the \textit{Further Work} section of reference \cite[p. 9]{lee2010detecting} Lee \textit{et al.} mention that they are working on a version of their algorithm which also expands all seeds in parallel.} 
\section{Algorithm}
We assume that each node is its own natural community $G$ at infinite resolution. The next vertex $V$ from the neighbourhood of $G$ included to $G$ is the one that increases the fitness of $G$ at the largest value of resolution denoted by $\alpha_\mathrm{incl}(G, V)$.

In pseudo code the growth of a natural community $G$  can be described as follows ($N(G)$ denotes the neighbourhood of $G$):
 \begin{algorithmic}[1]
\WHILE {$N(G)$ is not empty}
\FOR {each node $V$ in $N(G)$}  
 \STATE calculate $\alpha_\mathrm{incl}(G,  V)$ 
\ENDFOR
\STATE include the node with maximum $\alpha_\mathrm{incl}$ into $G$
\ENDWHILE
\end{algorithmic}
If two nodes have equal $\alpha_\mathrm{incl}$ MONC should include both (which we did not implement for the experiments described below). 

If we use the fitness function as defined by Lancichinetti \textit{et al.} \cite{lancichinetti2009detecting} a node cannot remain a single because for any alpha the module fitness of a single is always zero and the module fitness of two neighbours is always larger then zero. We can avoid this drawback of the algorithm by adding self-links to all nodes i.e. we assume that a node is a friend of itself or most similar to itself. To get results closer to those of reference \cite{lancichinetti2009detecting} we change the fitness function $F(G)$ only slightly by adding 1 to the numerator:
\begin{equation}
 f(G, \alpha) = \frac{k_\mathrm{in}(G) + 1}{(k_\mathrm{in}(G) + k_\mathrm{out}(G))^\alpha}.
\end{equation} 
From this definition we can derive a formula for calculating the maximum value of resolution $\alpha_\mathrm{incl}(G, V)$, where a node $V$ does not diminish the fitness of a module $G$ when included in it by demanding that for $\alpha < \alpha_\mathrm{incl}(G, V)$ we have $f(G \cup V, \alpha) > f(G, \alpha)$:\footnote{cf. \textit{Supplementary Information}}
\begin{equation}
 \alpha_\mathrm{incl}(G, V)  =  \frac{\log( k_\mathrm{in}(G \cup V) + 1 ) - \log (k_\mathrm{in}(G) + 1)}  {\log k_\mathrm{tot}(G \cup V) - \log k_\mathrm{tot}(G) },
\end{equation} 
where $k_\mathrm{tot}=k_\mathrm{in}+k_\mathrm{out}$ denotes the sum of the degrees of all nodes of a module.

We can calculate $k_\mathrm{in}(G \cup V)$ from $k_\mathrm{in}(G)$ and $k_\mathrm{tot}(G \cup V)$ from $k_\mathrm{tot}(G)$ i.e. the current values of the module from the preceding ones (which saves computing time). For this we define the interaction of a module and a node as    
\begin{equation}
 k_\mathrm{inter} (G, V) = \sum_{i \in G}A_{Vi},
\end{equation} 
where $A$ denotes the adjacency matrix of the undirected (and in general) weighted graph and calculate the degree of a node or its weight as the sum of the weights of its edges  \begin{equation}
A_{V+} = \sum_{i}A_{Vi}.   \end{equation} 

The weight of edges of internal nodes $k_\mathrm{in}$ is increased by $2\cdot k_\mathrm{inter}$ because both directions have to be taken into account: 
\begin{equation}
k_\mathrm{in}(G \cup V) = k_\mathrm{in}(G) + 2\cdot k_\mathrm{inter} (G, V).
\end{equation} 

The total of all weights is increased by the weights of the edges of the new node:
\begin{equation}
 k_\mathrm{tot}(G \cup V) = k_\mathrm{tot}(G) + A_{V+}.
\end{equation} 

We first include the neighbour $V$ of each node that improves the community's fitness at highest resolution. Then we continue with the new neighbourhood of $G\cup V$ until all nodes are included in the natural community. After each step we compare the current communities of all nodes to find duplicates. Thus we can reduce the number of communities treated by the inclusion algorithm and save further computing time. We merge overlapping natural communities of nodes. 

In addition to the changed fitness function described above, we deviated from LFK's approach in two more points. First, we do not allow the removal of nodes from a natural community. 
The LFK algorithm rechecks the fitness contribution of all community nodes after a new node has been added and excludes nodes if their removal increases the fitness. 
However, this possibility of exclusion contradicts the principle of locality. It can even lead to the exclusion of a seed node from its own natural community. In our networks of papers, removing nodes that reduce the fitness of a grown community is equivalent to shifting from the individual thematic perspective of the seed paper to a collective perspective of all papers in the community. 
Therefore our algorithm does not remove nodes from a community. Similarly,  Lee, Reid, McDaid, and Hurley \cite{lee2010detecting} implemented the 
LFK algorithm without exclusion mechanism. 

Another modification concerns the starting point of the algorithm. If a graph is characterised by a strong variation of its local density and the seed node is located in a high density region, the MONC algorithm immediately leaves this region because it searches for nodes with low degree first. These outside nodes only moderately increase the number of links leaving the community and thus often provide the earliest increase in fitness. We surmise that the LFK algorithm `repairs' this unwanted behaviour by allowing the exclusion of nodes with negative fitness. Since we suppressed the exclusion of nodes, 
we solved this problem by starting from cliques (i.e. totally linked subgraphs) instead of single nodes. Lee \textit{et al.}, who applied the LFK algorithm without the exclusion mechanism, also found that cliques as seeds
gave better results than single nodes \cite{lee2010detecting}.

While Lee \textit{et al.}\cite{lee2010detecting} use maximal cliques (i.e. cliques which are not subgraphs of other cliques), we optimise clique size by excluding nodes that are only weakly integrated. Thus, for our starting points we apply an analogon of the LFK exclusion mechanism. In detail, we exclude the node $V$ that diminishes the module fitness at lowest resolution, i.e. has the weakest coupling to the rest of the module $G$. Analogously to $\alpha_\mathrm{incl}$ we calculate $\alpha_\mathrm{excl}$ with
\begin{equation}
 \alpha_\mathrm{excl}(G, V)  =  \frac{\log( k_\mathrm{in}(G) + 1 ) - \log (k_\mathrm{in}(G \setminus V) + 1)}  {\log k_\mathrm{tot}(G) - \log k_\mathrm{tot}(G \setminus V) }.
\end{equation} 
This procedure is repeated until only two nodes remain in each clique.
From the set of shrinking cliques we select the one which is most resistant to further reduction i.e. those with highest $\alpha_\mathrm{excl}$ of the next node to be excluded. After its exclusion the rest of the clique would be less strongly coupled (for details see section \textit{Experiments} and cf. Figure \ref{fig_melting-h-clique} in \textit{Supplementary Information}). That means, we choose the most cohesive subgraph of a clique as optimal. 

After optimising all cliques larger than pairs we determine the optimal clique belonging to a seed node by searching for the clique where the seed is member and has its maximum $\alpha_\mathrm{excl}$. Nodes which are not member of any optimal clique remain single seeds. Every other node is  assigned to one clique, some of them to the same one.
%\newpage
\section{Data}
To compare our algorithm to that of Lancichinetti \textit{et al.} we first applied both to the network of social relations of 34 members of the well-known karate club observed by Zachary \cite{zachary1977information}. As Lancichinetti \textit{et al.} \cite{lancichinetti2009detecting} we used the unweighted version of this network.\footnote{s. \url{http://networkx.lanl.gov/examples/graph/karate_club.html}}

We also applied random LFK and MONC to a network of about 500 papers in volume 2008 of six information-science journals with a high portion of bibliometrics (see details in \textit{Supplementary Information}). 

In the network of information-science papers, two nodes (papers) are linked if they both have at least one cited source in common. The number of shared sources, which is normalised in order to account for different lengths of reference lists, provides a measure of the thematic similarity of papers. We start from the affiliation matrix $M$ of the bipartite network of papers and their cited sources. To account for different lengths of reference lists we normalise the paper vectors of $M$ to an Euclidean length of one. Then the element $a_{ij}$ of matrix $A = MM^\mathrm{T}$ equals Salton's cosine index of bibliographic coupling between paper $i$ and $j$. The symmetric adjacency matrix $A$ describes a weighted undirected network of bibliographically coupled papers. The elements of the main diagonal all equal 1, which means that a document is most similar to itself. We could proceed with this main diagonal i.e. with self-links but we omit them in the experiments described here (cf. \textit{Algorithm} section).

The main component of the bibliographic-coupling network of information science 2008 contains 492 papers. Two small components (three and two papers, respectively) and 34 isolated papers are of no interest for our experiments.

\section{Experiments}
%\newpage%\newpage
%\pagebreak
\subsection{Karate Club}
\begin{figure}[!t]
\centering
\includegraphics[width=3.5in]{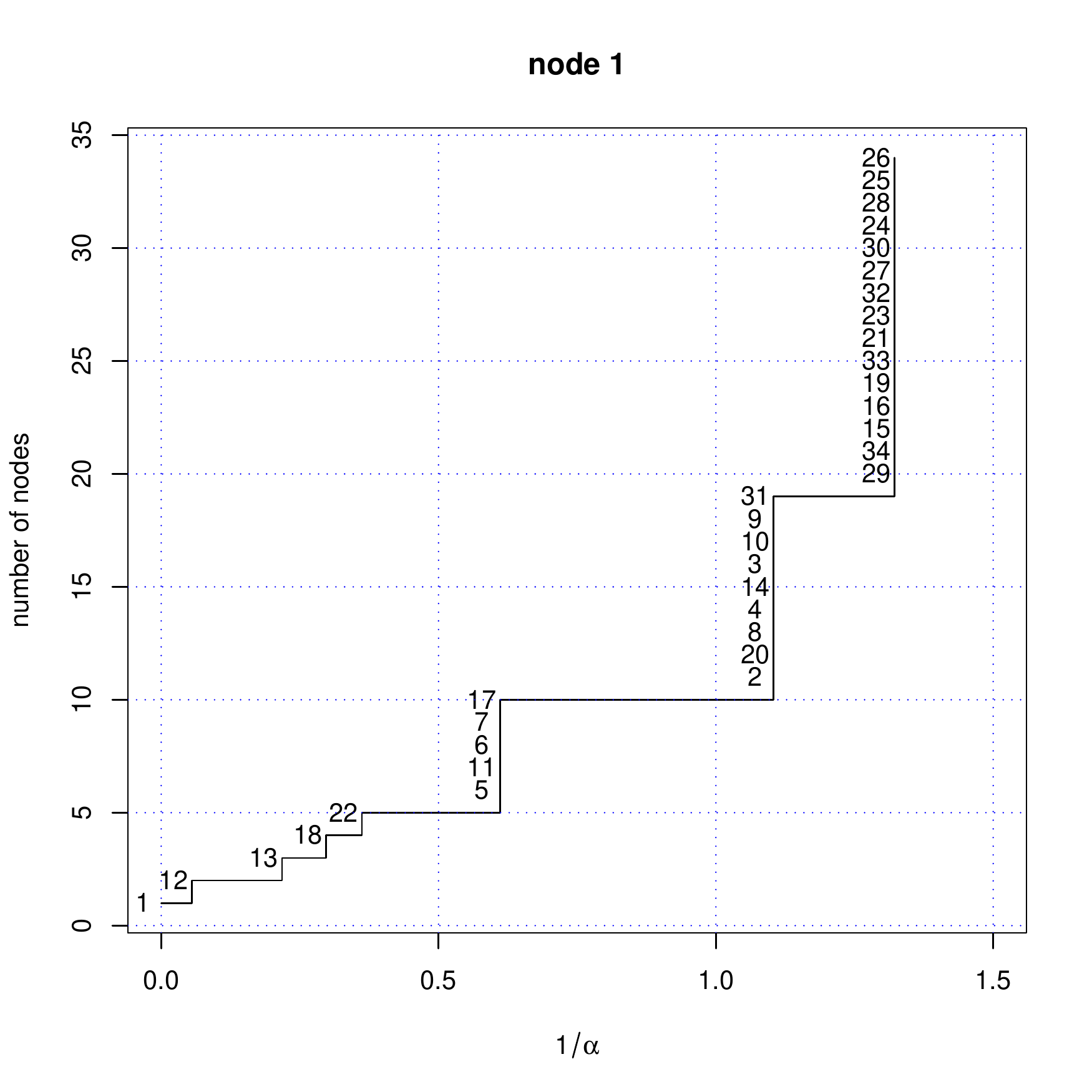}
\caption{Growing natural community of node 1 of Karate Club}
\label{fig_karate.comm.1}
\end{figure}

\begin{figure}[!b]
\centering
\includegraphics[width=3.5in]{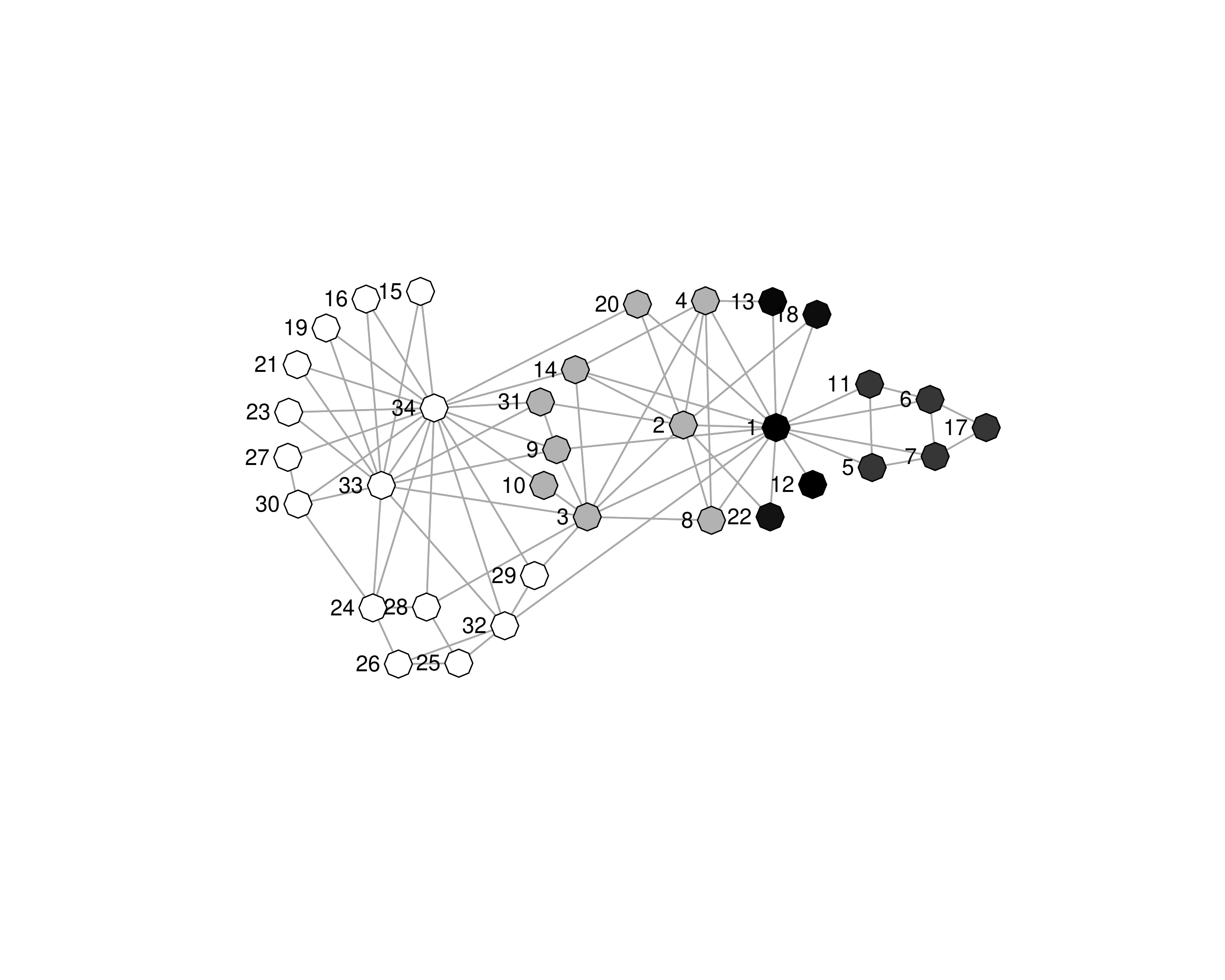}
\caption{Graph of growing natural community of node 1 of Karate Club}
\label{fig_karate.graph.comm.1}
\end{figure}

\begin{figure}[!t]
\centering
\includegraphics[width=3.5in]{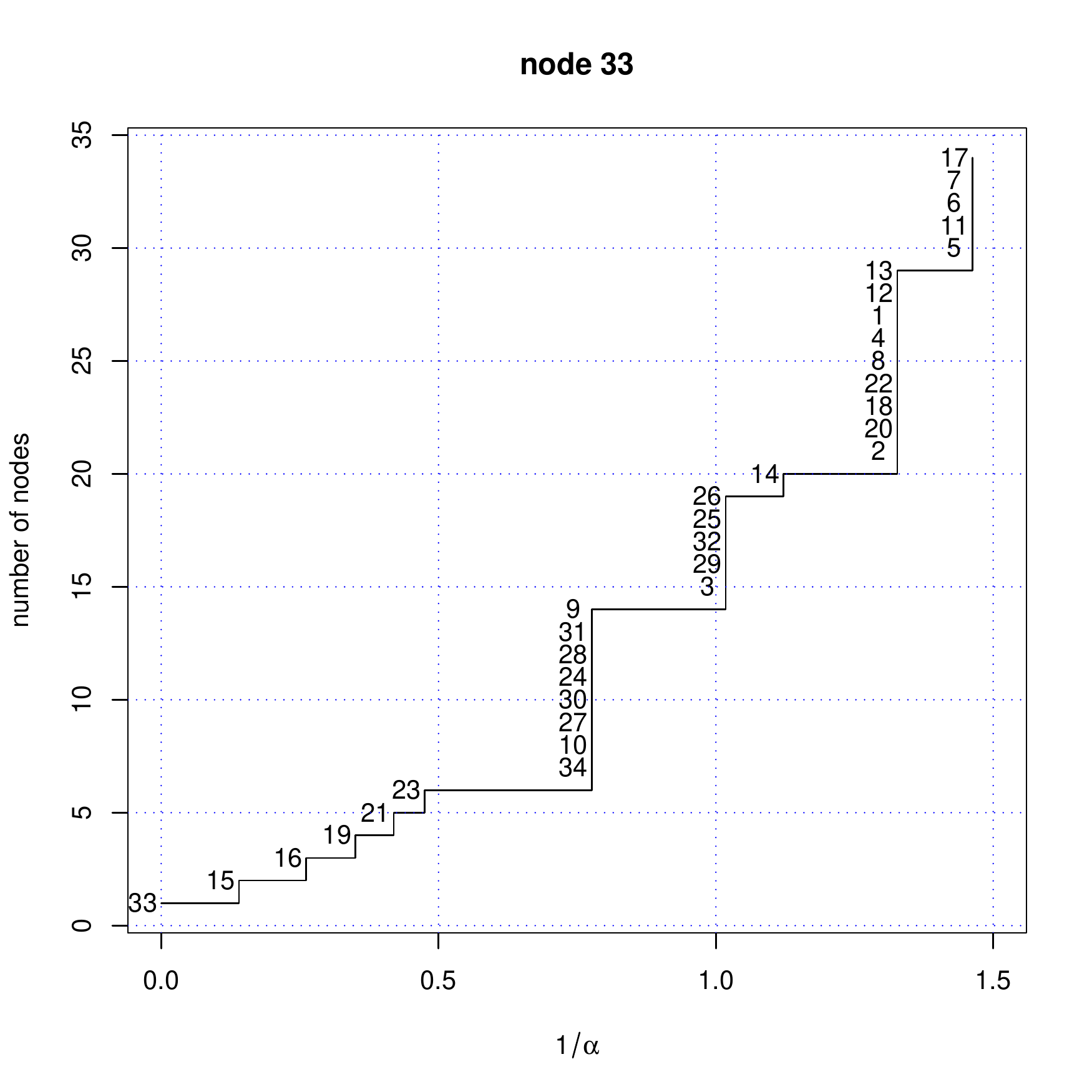}
\caption{Growing natural community of node 33 of Karate Club}
\label{fig_comm.33}
\end{figure}

\begin{figure}[!b]
\centering
\includegraphics[width=3.5in]{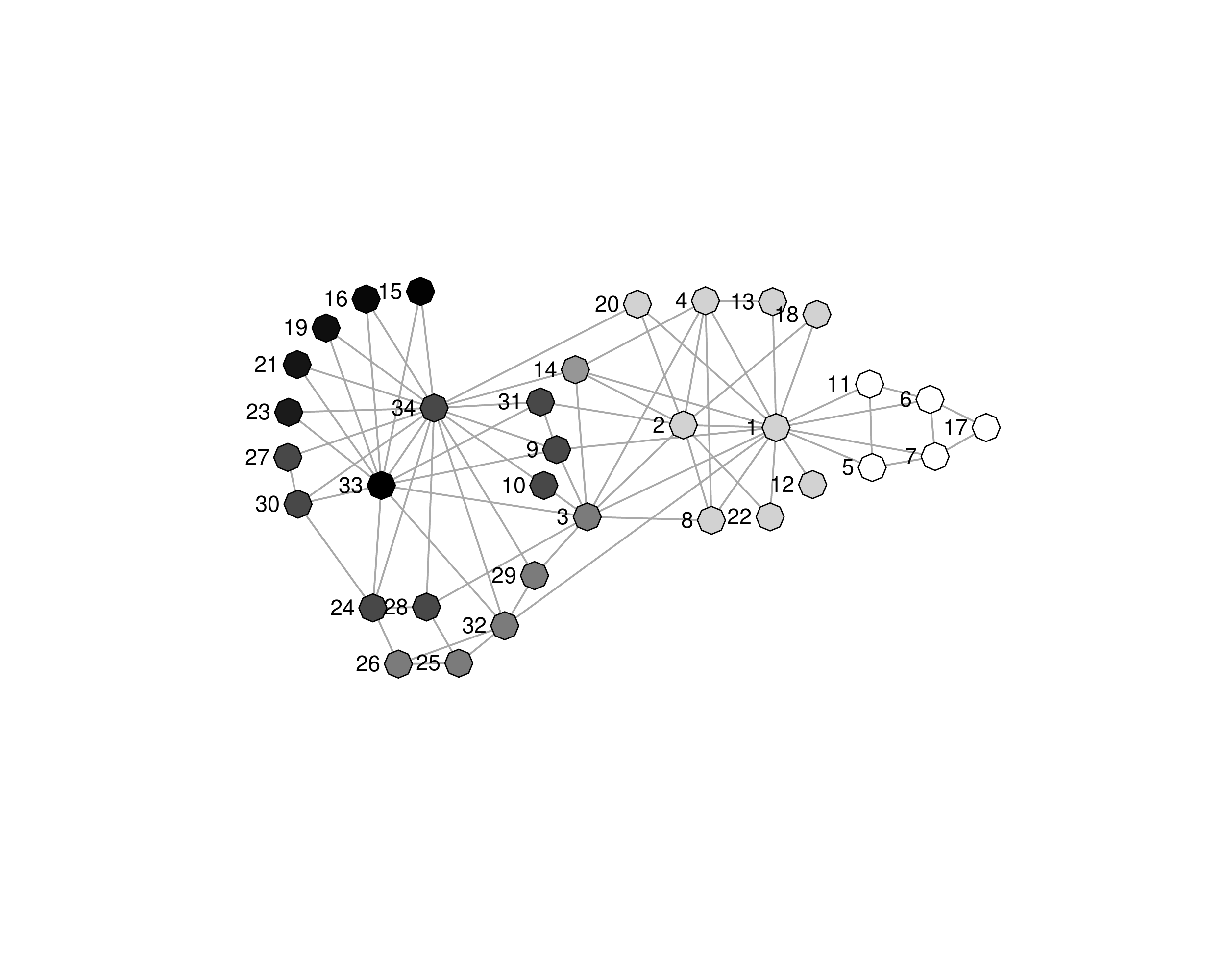}
\caption{Graph of growing natural community of node 33 of Karate Club}
\label{fig_graph.comm.33}
\end{figure}

\begin{figure}[!t]
\centering
\includegraphics[width=3.5in]{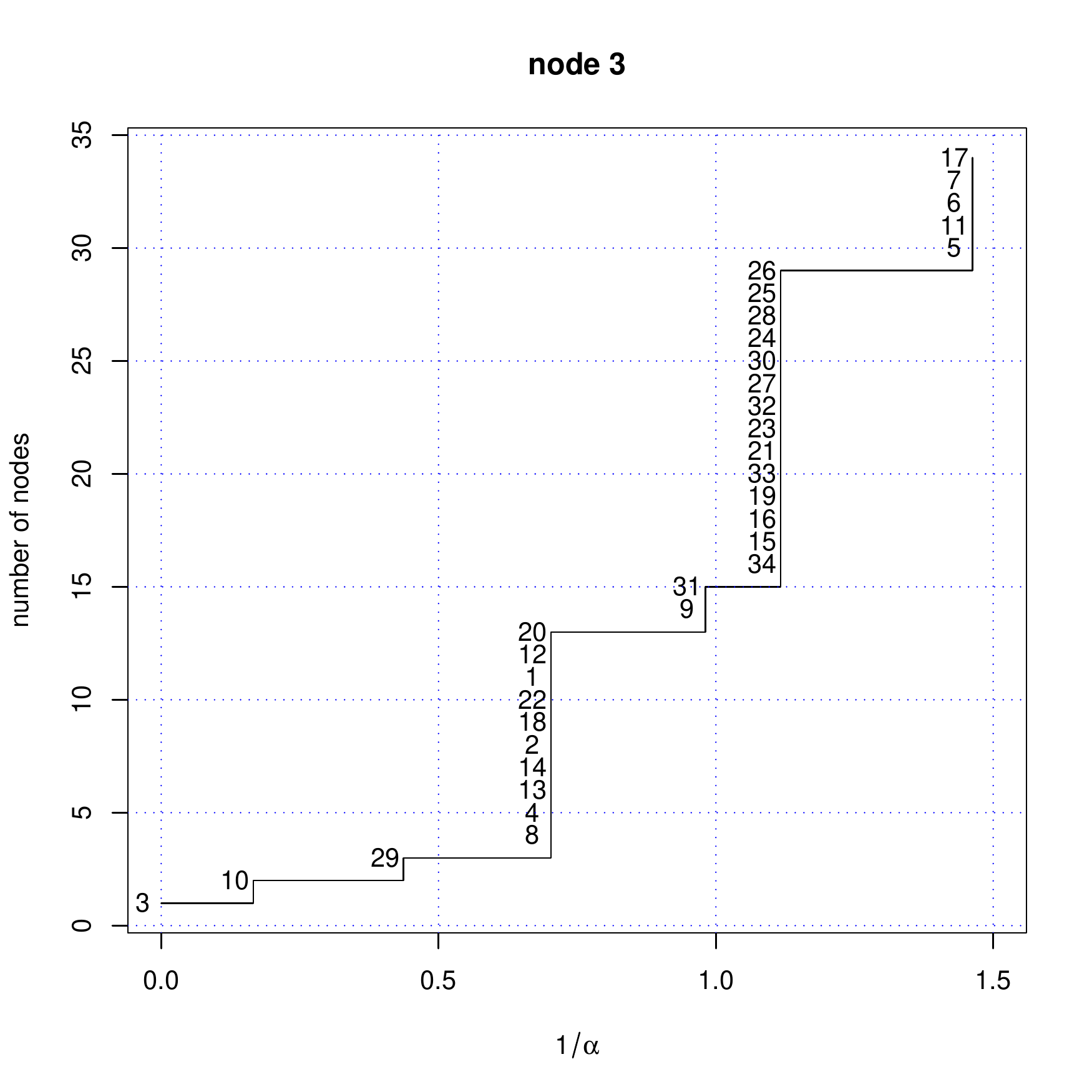}
\caption{Growing natural community of node 3 of Karate Club}
\label{fig_comm.3}
\end{figure}

\begin{figure}[!b]
\centering
\includegraphics[width=3.5in]{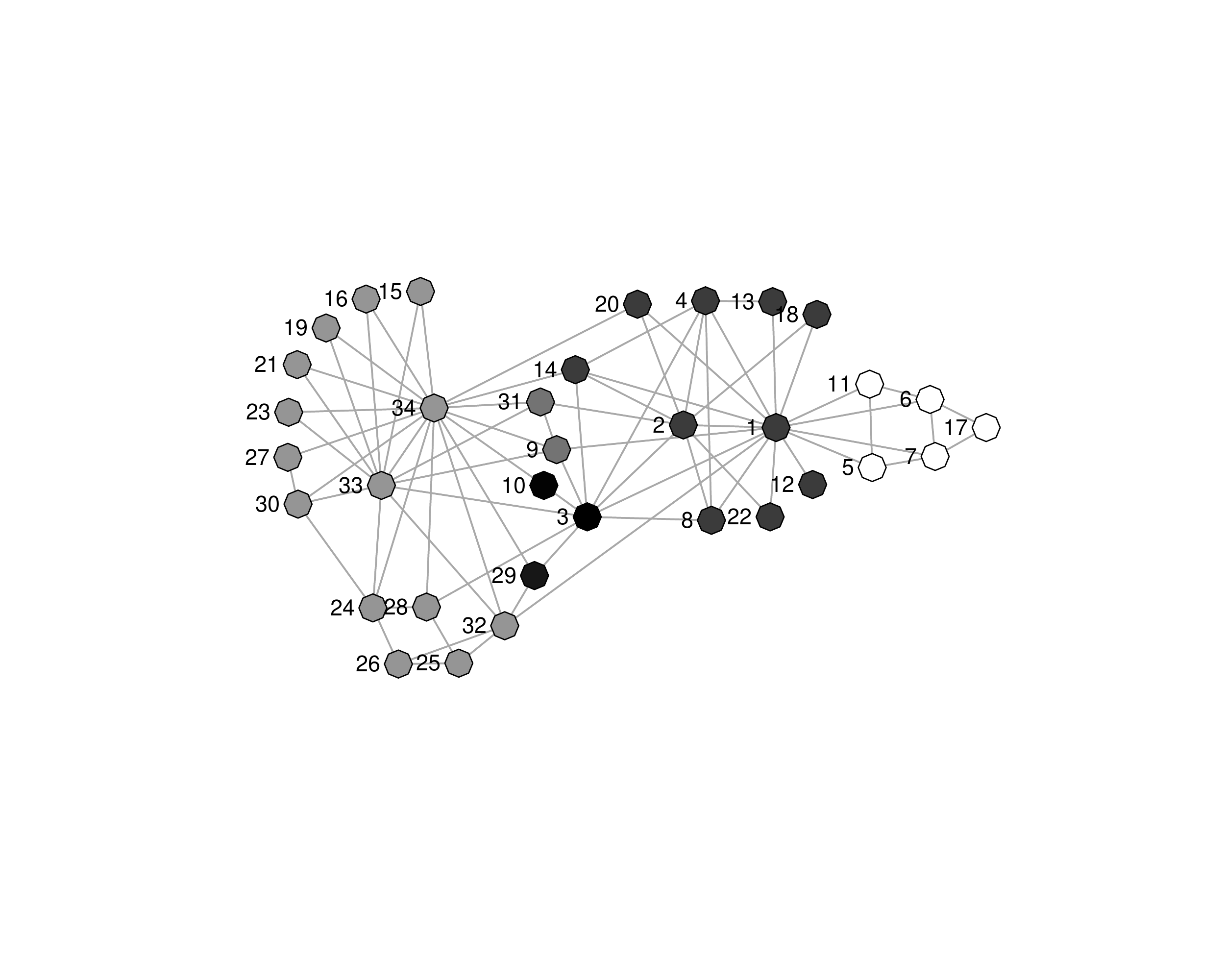}
\caption{Graph of growing natural community of node 3 of Karate Club}
\label{fig_graph.comm.3}
\end{figure}

Since the network of 34 karate club members is sparse---there is no clique with six or more fighters---we can apply MONC by starting from each node rather than using seed cliques. Figures \ref{fig_karate.comm.1}--\ref{fig_graph.comm.3} show the growing natural communities of three nodes. The step curve in the diagrams gives the growing number of nodes in the community as a function of $1/\alpha$. Each node is its own community at $1/\alpha = 0$. In our approach, the resolution always decreases, i.e. $1/\alpha$ cannot decrease. 

For example (cf. Figure \ref{fig_karate.comm.1}), even if nodes 11, 6, 7, and 17 enter the community of node 1 at lower $1/\alpha$ than their predecessor node 5, we display the same value of $1/\alpha$ for all five nodes because the higher resolution for the other four nodes becomes possible only after node 5 has been included. In other words, adding node 5 to the community changes the latter's properties in a way that would enable adding other nodes at a smaller value of $1/\alpha$.

The network graphs (Figures \ref{fig_karate.graph.comm.1}  and \ref{fig_graph.comm.33}) visualise the growth of communities by displaying the seed node in black, the last nodes joining in white, and the intermediate nodes on a grey scale corresponding to the resolution at which they come in. Lancichinetti \textit{et al.} \cite[Fig. 6(a), p. 10]{lancichinetti2009detecting} display the cover of the karate network they obtain in the resolution interval $.76 < \alpha < .84$ (which roughly equals the inverse resolution interval $1.2 < 1/\alpha < 1.3$). We see from the diagrams and graphs of nodes 1 and 33 that the MONC algorithm detects exactly the same cover in this interval, i.e. the same set of overlapping communities which cover the whole graph. 

Another cover in this resolution range is less  frequently obtained using the random LFK algorithm. It becomes visible in the diagram and graph of node 3, a node in the overlap of the two communities of the cover displayed by Lancichinetti \textit{et al.} In this resolution range the community of node 3 contains all nodes except the five nodes on the right end of the karate graph. The communities of these five nodes are identical and contain no other node in the resolution interval considered.

These examples indicate that for the karate club our MONC algorithm gives at least approximately the same results as the LFK algorithm. A detailed comparison reveals that 31 of the modules we found with our implementation of random LFK were also detected by MONC. Table \ref{table.Karate.modules} in \textit{Supplementary Information} lists the corresponding resolution intervals for both algorithms. Small differences are partly due to the different fitness functions (s. \textit{Algorithm} section) and partly due to the randomness of LFK. Further 22 LFK modules were not found by MONC. Their resolution intervals are mostly small~(maximum .2716, median .045). 

In addition MONC detected 23 modules which random LFK did not find. Each of these modules is found as an intermediate state of a growing natural community of only one seed node~(cf. column \textit{number of seeds} in Table \ref{table.Karate.modules} for the number of seed nodes of modules). 14 of them have $\alpha_{min} > 2$ and could not be found by LFK because in our implementation it run down from $\alpha = 2$ to $\alpha =0.65$. 

In summary, both algorithms are not equivalent but display similarities in many of their results. The LFK algorithm finds some modules MONC does not find. This is probably due to the exclusion mechanism of LFK that allows shifting a module away from its seed node.
%\newpage\newpage
\subsection{Information Science}
The 1812 maximal cliques of 492 bibliographically coupled information-science papers published in 2008 differ strongly in size. There are many small maximal cliques and some large ones. The density variation across the graph requires starting the MONC algorithm with seed cliques. 

The largest clique is formed by 46 papers which all cite the paper by J. E. Hirsch in 2005 where he proposes the \textit{h}-index: the Hirsch paper couples all these 46 papers. Many \textit{h}-clique papers also have the term \textit{h-index} in their titles but some of them discuss it only as a method among others. We reduce the \textit{h}-clique by the method described above to 21 papers which all have the \textit{h}-index or its derivatives as a central topic (cf. Figure \ref{fig_melting-h-clique} in \textit{Supplementary Information}; the distribution of clique sizes before and after reduction is given by Table \ref{table.InfSci.cliques} in \textit{Supplementary Information} section).

Most papers belong to more than one reduced clique. Each paper is assigned to the clique where it has its maximum $\alpha_\mathrm{excl}$. This leads to the selection of
357 reduced cliques. 16 papers have their highest $\alpha_\mathrm{excl}$ in the \textit{h}-clique. 275 cliques belong only to one node.  Three papers do not belong to any reduced clique and are therefore used as single paper seeds.

\begin{figure}[!t]
\centering
\includegraphics[width=3.5in]{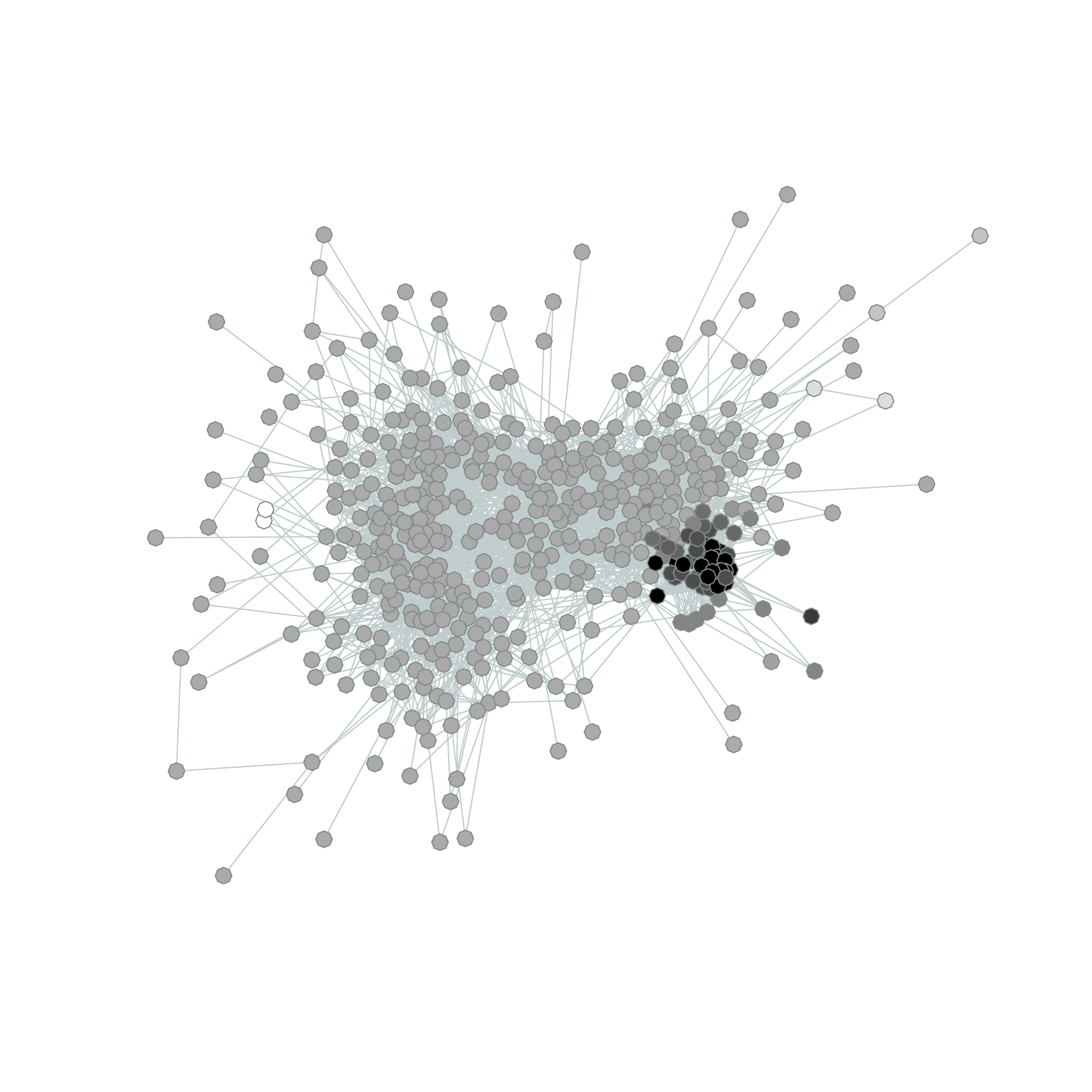}
\caption{Graph of growing natural community of \textit{h}-index clique (nodes are positioned by force directed placement)}
\label{fig_graph.comm.21}
\end{figure}

\begin{figure}[!b]
\centering
\includegraphics[width=3.5in]{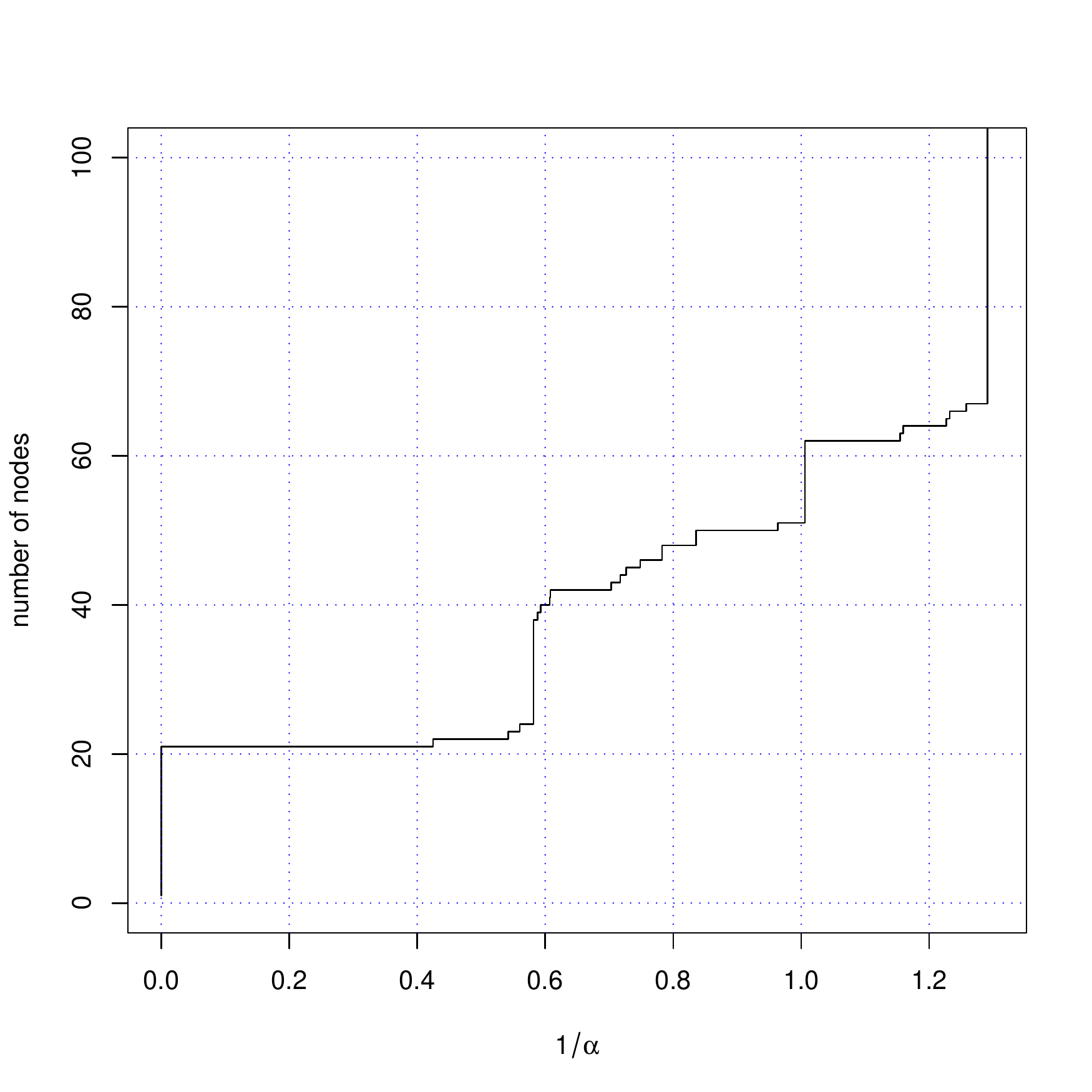}
\caption{Growing natural community of \textit{h}-index clique up to 100 papers}
\label{fig_comm.21}
\end{figure}

\begin{figure}[!t]
\centering
\includegraphics[width=3.5in]{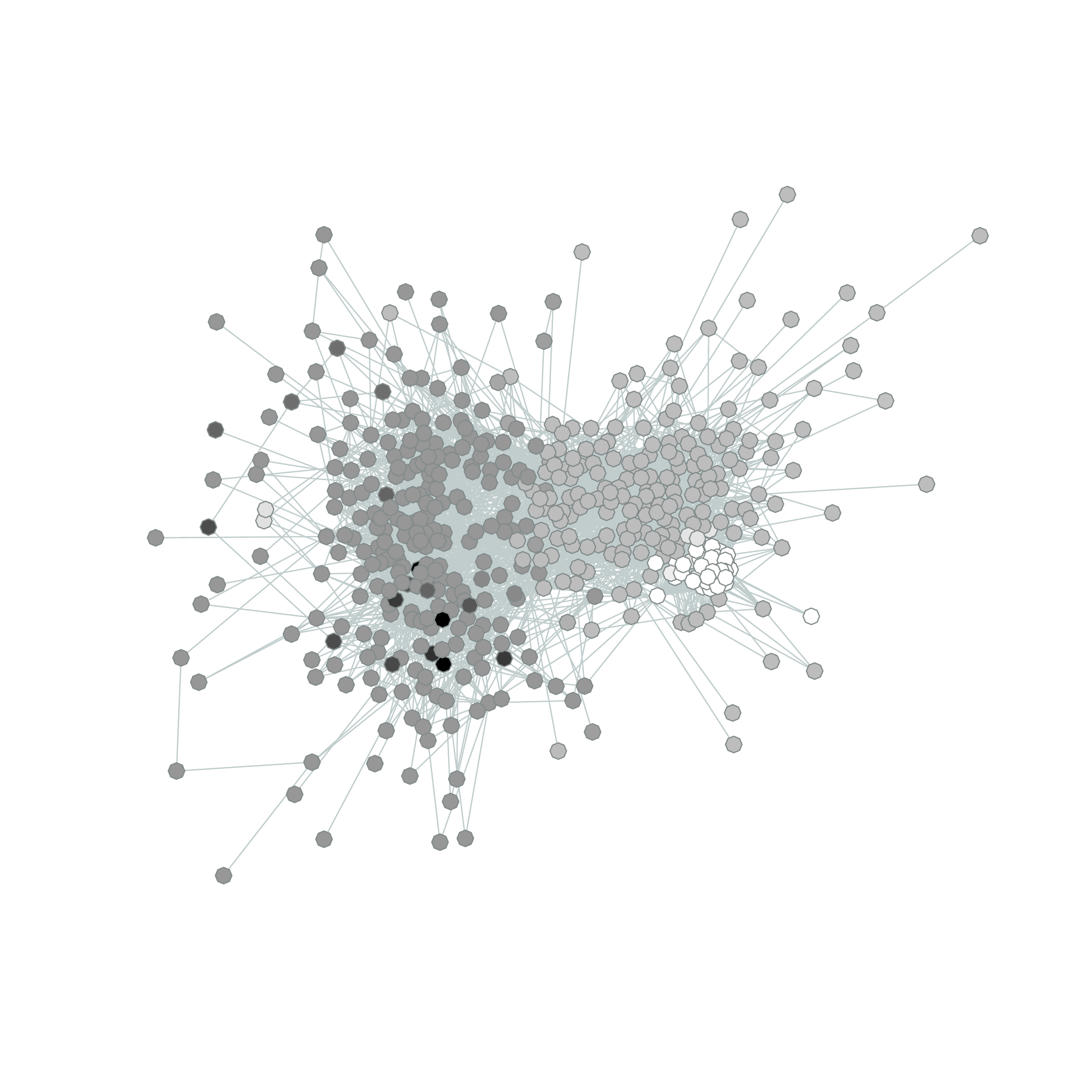}
\caption{Graph of growing natural community of information-retrieval papers}
\label{fig_graph.comm.1}
\end{figure}

\begin{figure}[!b]
\centering
\includegraphics[width=3.5in]{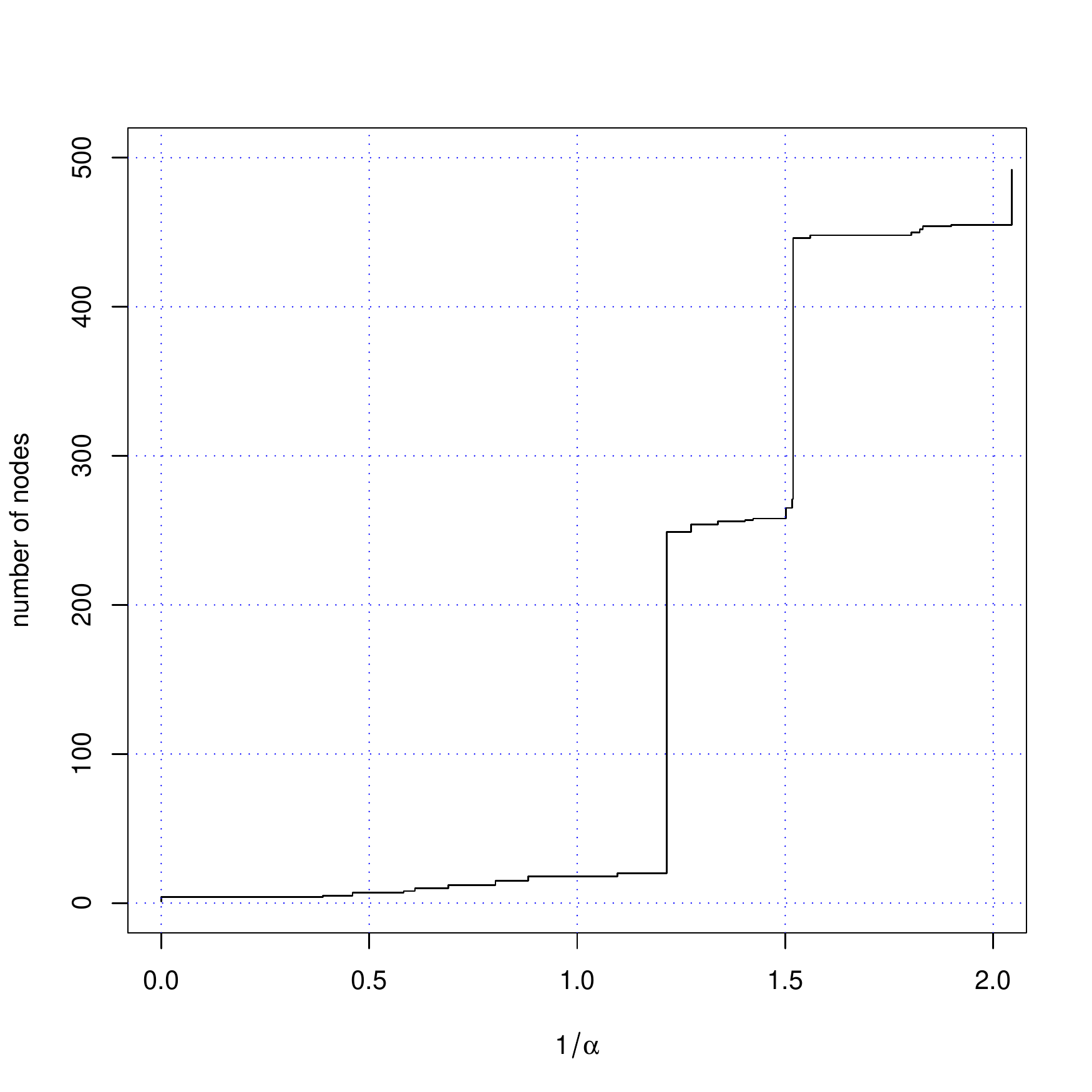}
\caption{Growing natural community of IR-papers}
\label{fig_comm.1}
\end{figure}

As an example, Figure \ref{fig_graph.comm.21} shows the graph of the growing natural community of one paper that has its highest $\alpha_\mathrm{excl}$ in the reduced \textit{h}-clique, whose 21 papers form the black core of the dark cloud in the figure. The corresponding diagram in Figure  \ref{fig_comm.21} visualises the growing natural community up to 100 papers. After collecting further 21 papers more or less related to the topic (mostly citing the Hirsch paper) the community's growth decelerates. This slow development lasts till $1/\alpha \approx 1$ ending up 
with 51 papers. We get the same succession of modules accumulating papers attached to the \textit{h}-community by applying the random LFK algorithm to information-science papers published in 2008. Even the corresponding thresholds of $\alpha$ obtained by both algorithms are nearly the same (Table \ref{table.h.modules} in \textit{Supplementary Information} section). Small differences between thresholds can be explained. First, MONC values are more precise because the LFK experiment was done in $\alpha$ steps of 1/100. Second, the MONC experiment is based on the modified fitness formula~(with  + 1 in the numerator). 

Figure \ref{fig_graph.comm.1} shows a sequential graph displaying intermediate steps while growing a community around a clique of information-retrieval (IR) papers (cf. Figure \ref{fig_comm.1}). It visualises the separation of IR papers (left) from papers in bibliometrics (right hand side).

MONC detected 5091 different modules as intermediate states of growing natural communities of nodes. Random LFK identified 1116 modules between $\alpha = 2$ and $\alpha = 0.1$  (in steps of 1/100). The $\alpha$ intervals of 3219 MONC modules overlap with this $\alpha$ region and are larger or equal to 0.2. The corresponding modules have therefore a realistic chance to be found by LFK, too. All in all, 211 modules across the whole spectrum of sizes were detected by both algorithms. LFK probably finds modules not found by MONC due to its exclusion mechanism. In addition, some smaller modules cannot be found by MONC because it here starts from cliques. MONC probably detects modules not found by LFK due to the latter's randomness.

The random LFK experiment started from $\alpha = 2$ and went down in steps of 1/100 to $\alpha = 0.1$. We implemented both algorithms as \textbf{R}-scripts.\footnote{\textbf{R} is an interpreted language and runs slower than compiled implementations.} 
LFK reached  $\alpha = 0.83$ after four hours and fifty minutes.\footnote{Intel(R) Xeon(R) CPU X5550@2.67 GHz with 72 GB RAM installed} 
The next value 0.82 is minimum $\alpha$ of 70 modules and took the algorithm more than three hours. All in all our slow random LFK implementation as an \textbf{R}-script (without storing community parameters, see \textit{Algorithm} section) needed 41 hours.  

A straightforward implementation of our MONC algorithm (also without storing community parameters, see \textit{Algorithm} section) reduced computation time to about 10 hours. The optimised version of MONC (with storing community parameters and neighbourhoods)
needed less than 12 minutes for the network of 492 nodes. 
In addition, the resolution thresholds computed by the MONC algorithm are much more accurate and the hierarchy of modules is detected automatically by MONC.

\begin{figure}[!t]
\centering
\includegraphics[width=3.5in]{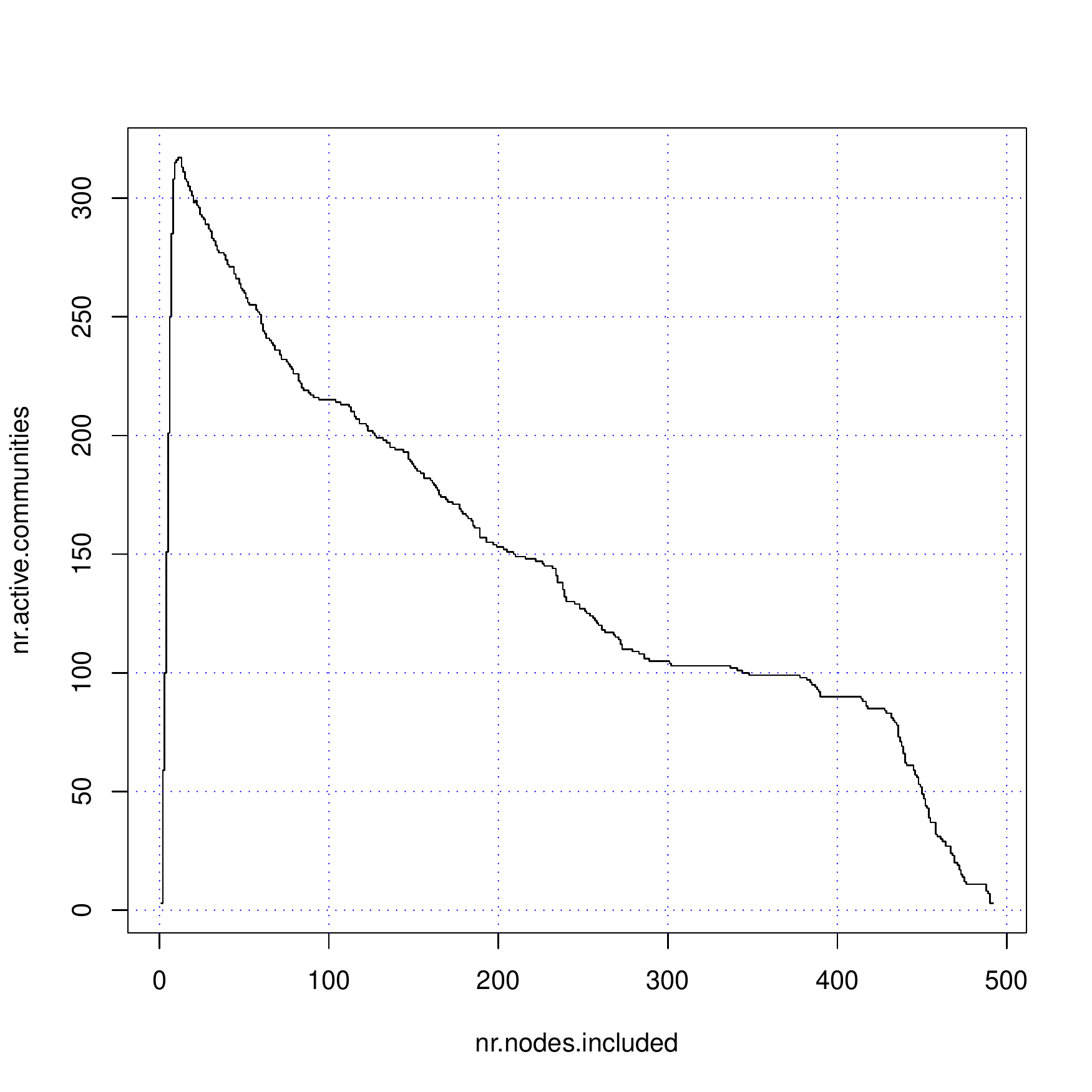}
\caption{Number of active communities in Information-Science experiment as a function of nodes included}
\label{fig_nr.active.comm}
\end{figure}

To illustrate merging of communities, Figure \ref{fig_nr.active.comm} displays how the number of active communities first rises to above 300~(of a maximum of 492 node communities or of 360 different seed cliques) and then is falling rapidly, thus making MONC faster. By active we denote growing communities which up to the current number of nodes included have not been made inactive by merging of communities. They will merge later. Only three communities survive before they are merged into the whole set of all 492 nodes.

\section{Summary and Conclusions}
The LFK algorithm detects overlapping natural communities of all nodes by maximising a local fitness function that enables the tuning of the procedure's resolution \cite{lancichinetti2009detecting}. Below some minimum resolution all nodes have the whole (connected) graph as their common natural community, while above some maximum resolution all nodes remain singles. If the algorithm is repeated for different resolution levels in an appropriate number of steps between maximum and minimum the hierarchical structure of the graph can be determined by comparing all communities found at all resolution levels considered. 

To maximise the local fitness function LFK includes nodes into a community that increase its fitness and excludes nodes reducing it. However, the exclusion of nodes violates the locality of the algorithm because nodes coming in later can `throw out' nodes that came in earlier, among them even the seed node. A variant of LFK without exclusion of nodes also gives reasonable results if it starts from maximal cliques instead of single nodes \cite{lee2010detecting}.

Another problem of the LFK algorithm is that it is time-consuming. LFK has been made faster by randomly choosing a new seed node that is not included  in any community detected so far \cite{lancichinetti2009detecting}. Diminishing the effects of randomness can and should be done by multiple runs at the same $\alpha$-level or by using small $\alpha$-steps. The random  procedure rests on the assumption that after each node is assigned to at least one community no further community has to be detected (cf. Lee \textit{et al}. \cite{lee2010detecting}, p. 3). If this assumption is unrealistic for the network considered the non-random LFK variant has to be applied.

We propose an algorithm (MONC) that also uses local fitness maximisation to include nodes but which is faster than LFK because it identifies overlapping natural communities of all nodes in one run. In our test on a weighted bibliometric graph of about 500 information-science papers non-optimised MONC was four times faster than our non-optimised random LFK implementation. Optimisation of MONC by storing community data accelerated it by a factor 50.

MONC includes nodes into communities but does not exclude nodes which diminish fitness. At each step MONC tests whether intermediate modules of growing communities of different nodes are equal. If this is the case, the two communities are merged. This not only makes MONC faster but automatically reveals the hierarchy of the network's modules that can be visualised as a dendrogram of overlapping communities. Thus, MONC can be seen as a truly hierarchical algorithm that clusters growing natural communities of a graph instead of its nodes.

If we follow the reasoning of Lancichinetti \textit{et al.} \cite[pp. 6--7]{lancichinetti2009detecting} we get $O(n^2 \log n)$ as the worst case complexity of random LFK algorithm. One factor $n$ is due to the exclusion mechanism and $log(n)$ is the order of the number of $\alpha$-levels needed to reveal the hierarchy of the network with $n$ nodes.  Hence, the computing time of non-random LFK variants should scale with $n^3 \log n$ and that of MONC with $n^2$ because MONC does not exclude nodes and uncovers the whole hierarchy in one run. Furthermore, MONC saves time due to  merging of communities. The estimation of complexity should be examined by applying MONC to benchmark graphs.

For each node MONC calculates the resolution thresholds at which its natural community grows by including new nodes from the neighbourhood, thereby identifying the (overlapping) natural communities of all nodes. Intervals of resolution at which the community does not expand are detected. These relatively stable intermediate modules of a community correspond to communities found in many LFK runs for different levels of resolution. MONC detects resolution intervals much more easily and more precisely than LFK. 

In the bibliometric test graph papers about the \textit{h}-index form an area that is very much denser than the rest of the graph because they constitute a clique (by citing the paper where Hirsch introduced the \textit{h}-index). Starting MONC with a node in a region of high density would (due to fitness maximization) immediately lead to sparse regions of the graph. We therefore use cliques as starting points. However, we do not use maximal cliques as Lee \textit{et al.} \cite{lee2010detecting} do because in bibliographic-coupling networks this could mean starting with papers that are only weakly related to the seed paper. We reduce the maximal cliques by excluding nodes until the maximum resolution threshold of the clique is obtained. This procedure  results in cliques with maximum cohesion as starting points of MONC. 

Some intermediate modules obtained by MONC while expanding communities for two test graphs coincide with~(often important) LFK communities and also exist for similar resolution intervals. We take this as a hint that MONC and LFK results are of comparable validity. By inspection, the structure of both test graphs obtained by MONC can be evaluated as reasonable and meaningful. This is why we expect MONC to produce valid modules  when applied to large benchmark graphs.

The local fitness function defined by Lancichinetti \textit{et al}.\cite{lancichinetti2009detecting} was selected by these authors among several alternatives (not specified by them) after some tests. We think that at least one alternative should be tested, namely the function
\begin{equation}
f(G, \beta) = \frac{k_{in}(G, \beta)}{k_{in}(G, \beta) + k_{out}(G)},                                                                      \end{equation} 
with $k_{in}(G, \beta) = k_{in}(G) + \beta |G|.$  That means that we calculate $k_{in}$ (the sum of internal degrees of nodes in $G$) but include self-links of weight $\beta$. Using weighted self-links for tuning resolution of modularity maximising methods was proposed and tested by Arenas, Fern{\'a}ndez, and G{\'o}mez in 2008\cite{arenas2008analysis}. They argue that the links between nodes are not changed by adding self-links. Thus the topology of the graph is not altered.

%The conclusion goes here.

\section*{Acknowledgement}
This work is part of a project in which we develop methods for measuring the diversity of research. The project is funded by the German Ministry for Education and Research (BMBF). We would like to thank all developers of \textbf{R}.\footnote{\url{http://www.r-project.org}}



\bibliographystyle{IEEEtran}
%\bibliography{Havemann2010local}
\bibliography{informetrics}
%\begin{thebibliography}{1}

%\bibitem{aa:lancichinetti2009detecting}
%\end{thebibliography}
%\pagebreak
\newpage
\section*{Supplementary Information}
To all tables in this section there is a reference in the main text. 

\begin{table}[!h]
%% increase table row spacing, adjust to taste
\renewcommand{\arraystretch}{1.3}
% if using array.sty, it might be a good idea to tweak the value of
% \extrarowheight as needed to properly center the text within the cells
\caption{533 Papers (528 articles and 5 letters) in volume 2008 of six information science journals (source: Web of Science)}
\label{table.journals}
\centering
%% Some packages, such as MDW tools, offer better commands for making tables
%% than the plain LaTeX2e tabular which is used here.
\begin{tabular}{rr}
\hline
journal                 &                papers  \\
\hline
INFORMATION PROCESSING \& MANAGEMENT     &    111 \\   
JOURNAL OF DOCUMENTATION	         &    40 \\ 
JOURNAL OF INFORMATION SCIENCE	         &    49 \\ 
JOURNAL OF INFORMETRICS                  &    31 \\ 
JOURNAL OF THE AMERICAN SOCIETY FOR      & \\ 
 INFORMATION SCIENCE AND TECHNOLOGY      &   176  \\ 
SCIENTOMETRICS	                         &   126 \\ 
\hline
sum                                      &   533 \\ 
\hline
\end{tabular}
\end{table}

\begin{table}[!b]
%% increase table row spacing, adjust to taste
\renewcommand{\arraystretch}{1.3}
% if using array.sty, it might be a good idea to tweak the value of
% \extrarowheight as needed to properly center the text within the cells
\caption{31 modules with at least two nodes in Karate-Club network found by MONC and by LFK (cf. section Results)}
\label{table.Karate.modules}
\centering
%% Some packages, such as MDW tools, offer better commands for making tables
%% than the plain LaTeX2e tabular which is used here.
\begin{tabular}{rr rrrr}
\hline
  number of        & MONC        &          &   number of     & LFK       & \\
nodes  &$\alpha_\mathrm{min}$ & $\alpha_\mathrm{max}$& seeds  &  $\alpha_\mathrm{min}$ & $\alpha_\mathrm{max}$ \\
\hline
34 & 0.0000000 & 0.7563793 & 34 & 0.6500 & 0.7662 \\
29  & 0.6835612 & 0.8952971 &  13 & 0.6887 & 0.8468 \\
20  &0.7535657 & 0.8915217 &  12 & 0.7630 & 0.9023 \\
19  &0.8915217 & 0.9823978 & 4 & 0.9024 & 1.1177 \\
19  &0.7563793 & 0.9056675 & 7 & 0.7663 & 0.8479 \\
14  &0.8332970 & 1.0117767 & 4 & 0.8480 & 1.0320 \\
14  &0.9823978 & 1.2892272 & 4 & 1.0000 & 1.2549 \\
12  &1.0117767 & 1.2542579 & 4 & 0.8650 & 1.2979 \\
12  &1.2892272 & 1.3175164 & 1 & 1.3186 & 1.3415 \\
11  &1.3175164 & 1.6524283 & 1 & 1.3524 & 1.3785 \\
 9  &1.8726915 & 1.9478173 & 1 & 1.9518 & 2.0000 \\
 6 &0.8119532 & 1.0716644 & 6 & 0.8663 & 1.1541 \\
 6  &1.2883392 & 2.1054487 & 1 & 1.3772 & 2.0000 \\
 5  &1.0716644 & 1.0928830 & 2 & 1.1551 & 1.2029 \\
 5  &0.6918777 & 1.0000000 & 5 & 0.7370 & 1.3569 \\
 5 &1.6367610 & 2.7625538 & 1 & 1.8201 & 2.0000 \\
 5  &2.1054487 & 2.3852809 & 1 & 1.6005 & 1.7242 \\
 4  &1.0928830 & 1.6040811 & 2 & 1.2075 & 1.3567 \\
 4  &0.8489011 & 1.1262455 & 4 & 0.9443 & 1.2892 \\
 4  &1.1262455 & 1.6204646 & 1 & 1.2893 & 1.9527 \\
 3  &1.4233850 & 2.2892242 & 1 & 1.7153 & 2.0000 \\ 
 3  &1.6204646 & 3.0578458 & 1  &  1.9528 & 2.0000 \\  
 3   &1.0503397 & 1.2598510 & 2 & 1.1664 & 1.7095 \\ 
 3  &1.1262455 & 2.7095113 & 3 & 1.2893 & 1.5849 \\ 
 3  &0.9578836 & 1.6586832 & 3 & 1.0966 & 2.0000 \\  
 2  &1.0000000 & 1.8690664 & 2 & 1.2969 & 2.0000 \\ 
 2  &1.2598510 & 3.8188417 & 1 & 1.3570 & 2.0000 \\ 
 2  &1.4321881 & 1.9631546 & 1 & 1.9434 & 2.0000 \\ 
 2  &1.0000000 & 1.5849625 & 2 & 1.3570 & 2.0000 \\ 
 2  &1.2223924 & 1.5849625 & 2 & 1.1446 & 2.0000 \\
 2  &0.8427577 & 2.7095113 & 2 & 1.1437 & 2.0000 \\

\hline
\end{tabular}
\end{table}

\begin{table}[!t]
%% increase table row spacing, adjust to taste
\renewcommand{\arraystretch}{1.3}
% if using array.sty, it might be a good idea to tweak the value of
% \extrarowheight as needed to properly center the text within the cells
\caption{Reducing cliques of 492 bibliographically coupled information-science papers 2008 ($S$ is original size, cf. section \textit{Experiments})}
\label{table.InfSci.cliques}
\centering
\begin{tabular}{lll lll}

  nr. of excluded nodes &&& && \\
\end{tabular}
\begin{tabular}{rrr rrr rrr rrr r}
$S$ &  0  & 1 & 2 & 3 & 4 & 5 & 6 & 7 & 8 &  13&   25 &  sum \\ 
\hline
  2 & 161 & 0 & 0 & 0 & 0 & 0 & 0 & 0 & 0 & 0 & 0 & 161 \\
  3 & 271 & 40 & 0 & 0 & 0 & 0 & 0 & 0 & 0 & 0 & 0 & 311 \\
  4 & 253 & 68 & 23 & 0 & 0 & 0 & 0 & 0 & 0 & 0 & 0 & 344 \\ 
  5 & 200 &115 & 38 & 24 & 0 & 0 & 0 & 0 & 0 & 0 & 0 & 377 \\
  6 & 147 & 91 & 40 & 15 & 8 & 0 & 0 & 0 & 0 & 0 & 0 & 301 \\
  7 & 54 & 52 & 25 & 18 & 3 & 0 & 0 & 0 & 0 & 0 & 0 & 152 \\ 
  8 & 22 & 29 & 20 & 8 & 7 & 1 & 1 & 0 & 0 & 0 & 0 & 88 \\ 
  9 &  8 & 5 & 10 & 5 & 2 & 1 & 0 & 1 & 0 & 0 & 0 & 32 \\   
  10 & 1 & 2 & 5 & 4 & 1 & 2 & 0 & 0 & 0 & 0 & 0 & 15 \\ 
  11 & 1 & 3 & 0 & 2 & 2 & 0 & 0 & 1 & 0 & 0 & 0 &  9 \\ 
  12 & 0 & 0 & 1 & 0 & 1 & 0 & 2 & 2 & 0 & 0 & 0 &  6 \\ 
  13 & 0 & 1 & 0 & 1 & 0 & 0 & 1 & 1 & 1 & 0 & 0 &  5 \\ 
  14 & 0 & 0 & 1 & 0 & 0 & 1 & 0 & 0 & 0 & 0 & 0 &  2\\
  15 & 0 & 0 & 0 & 0 & 1 & 0 & 0 & 0 & 1 & 0 & 0 &  2\\
  16 & 0 & 0 & 0 & 1 & 0 & 0 & 1 & 1 & 1 & 0 & 0 &  4\\
  18 & 0 & 0 & 1 & 0 & 0 & 0 & 0 & 0 & 0 & 0 & 0 &  1\\
  24 & 0 & 0 & 0 & 0 & 0 & 0 & 0 & 0 & 0 & 1 & 0 &  1\\
  46 & 0 & 0 & 0 & 0 & 0 & 0 & 0 & 0 & 0 & 0 & 1 &  1\\
$\Sigma$ & 1118 & 406 & 164 &  78 & 25 & 5 & 5 & 6 & 3 & 1 & 1 & 1812\\
\hline
\end{tabular}
\end{table}

\begin{figure}[!b]
\centering
\includegraphics[width=3.5in]{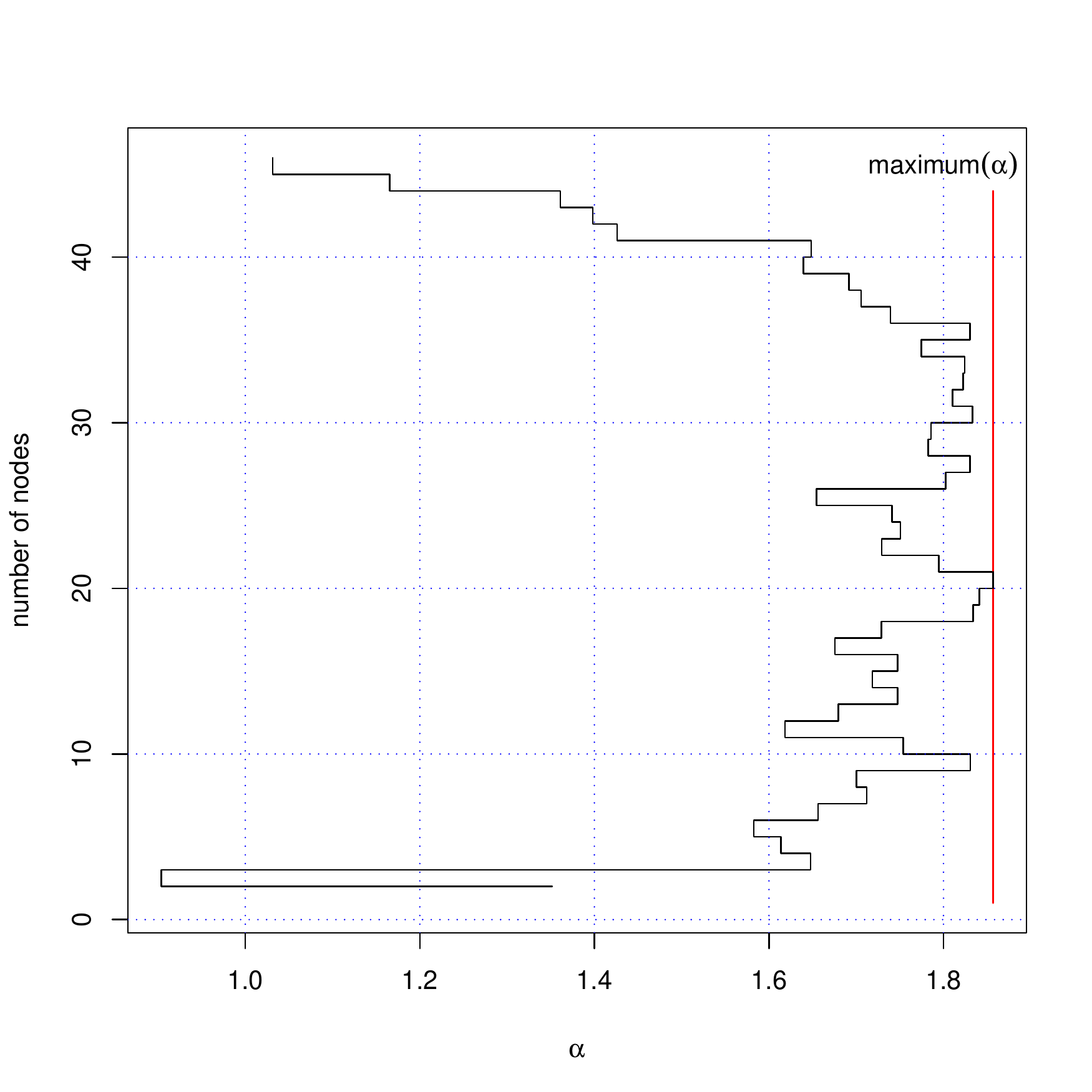}
\caption{Optimisation of a clique of 46 $h$-index papers to 21 core papers}
\label{fig_melting-h-clique}
\end{figure}

\begin{table}[!h]
%% increase table row spacing, adjust to taste
\renewcommand{\arraystretch}{1.3}
% if using array.sty, it might be a good idea to tweak the value of
% \extrarowheight as needed to properly center the text within the cells
\caption{Comparison of α thresholds obtained by random LFK and by MONC algorithm, respectively, in a succession of modules growing from h-clique}
\label{table.h.modules}
\centering
\begin{tabular}{lllll}
\hline
   &  random    &                    &  MONC& \\
      &LFK   &                    &   (rounded)&\\
     nr. nodes& $\alpha_\mathrm{min}$  & $\alpha_\mathrm{max}$  &  $\alpha_\mathrm{min}$ &$\alpha_\mathrm{max}$\\
\hline
        38 & 1.71 & 1.71	 &   1.7019 & 1.7353 \\
        39 & 1.70 & 1.70	 &   1.6880 & 1.7019\\
        40 & 1.66 & 1.69	 &   1.6361 & 1.6880\\
        42 & 1.43 & 1.65	 &   1.4233 & 1.6453\\
        43 & 1.40 & 1.42	 &   1.3955 & 1.4233\\
        44 & 1.39 & 1.39	 &   1.3587 & 1.3955\\
        45 & 1.35 & 1.38	 &   1.2817 & 1.3587\\
        46 & 1.29 & 1.34	 &   1.2792 & 1.2817\\
        48 & 1.21 & 1.28	 &   1.1903 & 1.3103\\
        50 & 1.05 & 1.20	 &   1.0308 & 1.1910\\		
        51 & 1.00 & 1.04	 &   0.9956 & 1.0308\\

\hline
\end{tabular}
\end{table}

\pagebreak
\newpage
Figure \ref{fig_melting-h-clique} illustrates the optimisation of maximal cliques by exclusion of nodes. Nodes with minimum $\alpha_\mathrm{excl}$ are excluded one after the other from the clique. From the set of shrinking cliques we select the one before maximum $\alpha_\mathrm{excl}$  (marked by the vertical line) is reached.

\newpage
We now derive the formula for calculating the maximum value of $\alpha$, where a node $V$ does not diminish the fitness of a module $G$ when included in it. For $V$ in neighbourhood of $G$ we demand therefore

\begin{equation}
f(G \cup V, \alpha) > f(G, \alpha).
\end{equation} 
With definitions given in \textit{Algorithm} section we then have
\begin{equation}
\frac{k_\mathrm{in}(G \cup V) + 1}{k_\mathrm{tot}(G \cup V)^{\alpha}} > 
\frac{k_\mathrm{in}(G) + 1}{k_\mathrm{tot}(G)^{\alpha}}
\end{equation}
and therefore 
\begin{equation}
\frac{k_\mathrm{in}(G \cup V) + 1}{k_\mathrm{in}(G) + 1} > 
\left[ \frac{k_\mathrm{tot}(G \cup V)}{k_\mathrm{tot}(G)} \right] ^{\alpha}.
\end{equation}
We take logarithm on both sides of this equation and get
\begin{equation}
\log\frac{k_\mathrm{in}(G \cup V) + 1}{k_\mathrm{in}(G) + 1} >
\alpha \log\frac{k_\mathrm{tot}(G \cup V)}{k_\mathrm{tot}(G)}.
\end{equation}
That means, if $\alpha < \alpha_\mathrm{incl}$ with 
\begin{equation}
\alpha_\mathrm{incl} = \frac{\log({k_\mathrm{in}(G \cup V) + 1)}-\log({k_\mathrm{in}(G) + 1)}}{\log{k_\mathrm{tot}(G \cup V)}-\log{k_\mathrm{tot}(G)}}
\end{equation}
we have $f(G \cup V, \alpha) > f(G, \alpha)$.

%\bibliography{informetrics}

\end{document}